\documentstyle[12pt,aaspp4]{article}

\lefthead{Bekki Kenji and   Yasuhiro Shioya}
\righthead{Photometric  evolution of mergers}

\begin{document}
\title{Photometric evolution of dusty starburst  mergers:
On the nature of ultra-luminous infrared galaxies.}

\author{Kenji Bekki} 
\affil{Division of Theoretical Astrophysics,
National Astronomical Observatory, Mitaka, Tokyo, 181-8588, Japan} 

\and

\author{Yasuhiro Shioya} 
\affil{
Astronomical Institute, 
Tohoku University, Sendai, 980-8578, Japan}

\begin{abstract}

By performing N-body simulations of chemodynamical evolution of galaxies
with dusty starbursts,
we investigate photometric evolution
of gas-rich major mergers
in order to explore the nature of ultraluminous infrared galaxies (ULIRGs)
with the total infrared luminosity 
($L_{\rm IR}$ 
for $8\sim 1000$ $\mu$m) of 
$\sim$ $10^{12}$ $L_{\odot}$.
Main results are the following five.

 (1) Global colors and absolute magnitudes the during dusty starburst
of a major merger
do not change with time  significantly,
 because interstellar dust heavily obscures
young starburst populations that could cause 
rapid evolution of photometric properties 
of the merger.

 (2) Dust extinction of stellar populations in a galaxy merger with
large infrared luminosity ($L_{\rm IR}$ $>$ $10^{11}$ $L_{\odot}$)
is selective in the sense that 
younger stellar populations are preferentially obscured by dust
than old ones.
This is because younger populations are located
in the central region where a larger amount of dusty
interstellar gas can be transferred from the outer gas-rich regions
of  the merger.

(3) Both $L_{\rm IR}$ and 
the ratio of $L_{\rm IR}$ to $B$ band luminosity
$(L_{\rm B}$) increases as the star formation
rate increase during the starburst of the present merger model,
resulting in the positive correlation between $L_{\rm IR}$ and 
$L_{\rm IR}/L_{\rm B}$. 
The dust temperature $T_{\rm dust}$ and  the flux ratio of
$f_{60 \mu \rm m}/f_{100 \mu \rm m}$ also 
increase with the increase of the star formation rate,
because a larger number of young stars
formed by massive starburst can heat the dusty interstellar gas
as the star formation becomes larger.

(4) The star formation efficiency, total gas mass, 
the degree of dust extinction ($A_{V}$), $T_{\rm dust}$, 
$L_{\rm IR}$, $L_{\rm IR}/L_{\rm B}$, and $f_{60 \mu \rm m}/f_{100 \mu \rm m}$
depend strongly
on the separation of two cores of the merger,
which clearly reflects the fact that dynamical processes of
galaxy merging play an important role in
determining  the photometric  evolution
of dusty starbursts in the merger.

(5) The two-dimensional distribution
of  global colors (e.g., $R-K$) shows  a negative color
gradient during starburst, 
mainly because  central  young starburst populations are preferentially
and very heavily obscured by dust. 
Furthermore,  the peak of the two dimensional
distribution of  $f_{60 \mu \rm m}$ 
coincides  with that of $A_{V}$.
These results clearly suggest that the spatial distribution
of gas and stars is one  important determinant of
the two-dimensional photometric properties of major mergers.

Our numerical  results clearly demonstrate that
dynamical processes of major galaxy  merging,
which can control the time evolution of the relative
spatial distribution of dusty interstellar gas
and young stars,
play a vital role in determining
photometric properties of ULIRGs.

\end{abstract}


\section{Introduction}

Ultraluminous infrared galaxies (ULIRGs) with $L_{\rm IR}(8-1000 \mu m)$
$\sim$ $10^{12}$ $L_{\odot}$
discovered by the $Infrared$ $Astronomical$ $Satellite$ (IRAS)
are considered to be ideal laboratories  for studying
the formation of dusty starburst galaxies, elliptical galaxies, 
and QSOs.
Detailed morphological studies of these ULIRGs
have revealed that virtually all show evidence of galaxy merging
or strong tidal interaction (e.g., tidal tails and double or multiple
nucleus) and thus that most of ULIRGs are formed by galaxy interaction/merging
(e.g., Sanders et al. 1988; Clements et al. 1996; Murphy et al. 1996;
Scoville et al. 1999; Surace et al. 1999).  
These ULIRGs are observed to show mixtures of two types of activity; Starburst
and  active galactic nuclei (AGN) (Sanders et al 1988; Sanders \& Mirabel 1996;
Veilleux, Kim, \& Sanders 1999; Zheng et al. 1999).
It is still controversial whether dust-enshrouded AGN or a nuclear
dusty starburst is the primary  energy source for the huge luminosity of ULIRGs
(Sanders \& Mirabel 1996 for a review; 
Genzel et al. 1998; Lutz et al. 1998; Smith, Lonsdale, \& Lonsdale 1998).
All of these of ULIRGs are rich in molecular gas with the total mass
of $\rm H_{2}$ ranging from $10^9$ $M_{\odot}$ to 3 $\times$ 
$10^{10}$ $M_{\odot}$ (Sanders et al. 1988), and a large fraction
of this gas is in very dense regions near nuclei (Solomon et al. 1992).
ULIRGs appear to be about 2 times more numerous than optically
selected QSOs and the luminosity function of ULIRGs is nearly similar
to that of QSOs (Soifer et al. 1986).
These results give rise to the argument that quasar-like
activity is also being initiated in major mergers
and ULIRGs can finally evolve into QSOs (Sanders et al. 1988).
Recent high-resolution optical/infrared imaging of ULIRGs
have revealed a number of very bright star-forming knots
probably formed by merging processes 
and furthermore clarified diverse morphology reflecting
their varying stages of galactic dynamical evolution (Scoville et al. 1999;
Surace et al. 1999).

These peculiar phenomena seen in ULIRGs (i.e., ongoing major mergers)
have raised theoretical interests in the interstellar gas and star formation
process in gas-rich major mergers with strong dusty starbursts.
Numerical simulations of galaxy mergers including gas dynamics
indicate that a large amount  of gas, as much as $10^9$ $M_{\odot}$,
can be funneled to the central kpc region of the remnant
(Barnes \& Hernquist 1991, 1992; Olson \& Kwan 1990; Noguchi 1991).
Mihos et al. (1992) suggested that the gas accumulated in the central
kpc region of galaxy mergers
can be converted into massive starburst, though the magnitude
of the starburst is not strong enough to explain that observed in
ULIRGs.  
By using a more sophisticated model of hydrodynamical
evolution of galaxy mergers, Mihos \& Hernquist (1996)  
demonstrated that the strength and the epoch of massive starburst
depend on initial internal structure of merger progenitor disks
rather than initial orbital configurations of the disks.
Bekki \& Noguchi (1994) and Bekki (1995) demonstrated that
the dynamical heating by two sinking cores 
and subsequent dissipative cloud-cloud collisions in a merger
can drive a large fraction of disk interstellar gas 
into the central tens pc where nuclear starburst and AGN
activity are observed for  ULIRGs.
Gerritsen \& Icke (1999) investigated
how thermal energy from supernovae (i.e., feedback
of supernovae) affects the dynamical evolution of interstellar
gas and consequently controls the star formation rate in a major
merger and found that the supernovae energy from the nuclear
high density 
gas region ($\sim$ 1 kpc) can prevent the star formation in the outer region
of the merger.  
Although these theoretical studies have been quite successful
in clarifying the detailed processes of gaseous inflow to
the central 10 $\sim$ 1000 pc and in describing star formation
history for gas-rich major galaxy mergers,
they have not yet investigated at all how photometric and spectroscopic
properties of merger evolve with time.
Therefore it remains highly uncertain  
how the photometric properties
of ULIRGs is determined by the detailed dynamical
processes of major galaxy merging.

The purpose of the present paper 
is to investigate numerically morphological,
structural, and photometric evolution
of gas-rich major mergers with dusty starburst in an
explicitly self-consistent manner.
By using a new numerical code for  calculating the spectral
energy distribution (SED) of a galaxy
with dusty interstellar gas (Bekki \& Shioya 2000a),
we particularly investigate the following five points:
(1) The fundamental properties  of SEDs of dusty starburst mergers
with  the infrared luminosity larger than 
$10^{11}$ $L_{\odot}$,
(2) time evolution of the relative spatial distribution
of dusty gas and old and new stars, 
(3) The dependence of $A_{V}$ on stellar ages and
radius from the center of mass of a merger, 
(4) time evolution of colors, absolute magnitude, surface brightness,
dust temperature,
and the degree of dust extinction ($A_{V}$), and
(5) physical correlations between $L_{\rm IR}$, central gas density,
luminosity ratio (e.g., $L_{\rm IR}/L_{B}$, where $L_{B}$
is the $B$ band luminosity), star formation efficiency,
and morphology during starburst.
The above points have not been investigated extensively by
previous theoretical studies. Accordingly, the present numerical
results can provide new clues to the nature and the origin
of ULIRGs.

The layout of this paper is as follows.
In \S    2, we summarize  numerical models used in the
present study and describe briefly  the methods for 
deriving the SEDs corrected by internal dust extinction.
In \S 3, we present numerical results on the time evolution
of morphology, SED, and photometric properties in a gas-rich
major merger. In \S 4, we discuss the nature of ULIRGs,
in particular, the importance of the relative distribution
of gas and stars in determining photometric and spectroscopic
properties of ULIRGs. 
The conclusions of the preset study
are  given in \S 5.

\section{Model}

Since  numerical methods  and  technique for solving chemodynamical
and photometric evolution of major galaxy mergers with dusty starburst
have been already given  in detail by Bekki \& Shioya (2000a),
we briefly describe the present merger model. 
 We construct  models of galaxy mergers between gas-rich 
 disk galaxies with equal mass by using the Fall-Efstathiou (1980) model.
 The total mass and the size of a progenitor disk are $M_{\rm d}$
 and $R_{\rm d}$, respectively. 
 From now on, all the mass and length are measured in units of
  $M_{\rm d}$ and  $R_{\rm d}$, respectively, unless specified. 
  Velocity and time are 
  measured in units of $v$ = $ (GM_{\rm d}/R_{\rm d})^{1/2}$ and
  $t_{\rm dyn}$ = $(R_{\rm d}^{3}/GM_{\rm d})^{1/2}$, respectively,
  where $G$ is the gravitational constant and assumed to be 1.0
  in the present study. 
  If we adopt $M_{\rm d}$ = 6.0 $\times$ $10^{10}$ $ \rm M_{\odot}$ and
  $R_{\rm d}$ = 17.5 kpc as a fiducial value, then $v$ = 1.21 $\times$
  $10^{2}$ km/s  and  $t_{\rm dyn}$ = 1.41 $\times$ $10^{8}$ yr,
  respectively.
  In the present model, the rotation curve becomes nearly flat
  at  0.35  radius with the maximum rotational velocity $v_{\rm m}$ = 1.8 in
  our units.
  The corresponding total mass $M_{\rm t}$ and halo mass $M_{\rm h}$
  are 5.0  and 4.0 in our units, respectively.
  The radial ($R$) and vertical ($Z$) density profile 
  of a  disk are  assumed to be
  proportional to $\exp (-R/R_{0}) $ with scale length $R_{0}$ = 0.2
  and to  ${\rm sech}^2 (Z/Z_{0})$ with scale length $Z_{0}$ = 0.04
  in our units,
  respectively.
 The central bulge component is represented by a Plummer sphere
with the mass of 0.25  
and the scale length of 0.04 in our units,
which  means  that the mass ratio of
bulge to disk is 0.25 and the scale length ratio of bulge to
disk is 0.2. 
  The collisional and dissipative nature 
  of the interstellar medium is  modeled by the sticky particle method
  (\cite{sch81})
and the initial fraction of gas mass ($f_{g}$) in
a disk is set to be 0.2.
  The  number of particles 
  for an above  isolated galaxy is 
  10000 for dark halo,   
3165 for stellar bulge,
10000 for stellar disk components, 
and 10000 for gaseous ones.

    In all of the simulations of  mergers, the orbit of the two disks is set to be
    initially in the $xy$ plane and the distance between
    the center of mass of the two disks ($r_{\rm in}$)
  is  140 kpc. 
The pericenter
distance ($r_{\rm p}$)  and the orbital eccentricity ($e$)
are assumed to be 17.5 kpc and 1 (i.e., parabolic), respectively. 
    The spin of each galaxy in a   merger
is specified by two angle $\theta_{i}$ and
    $\phi_{i}$, where suffix  $i$ is used to identify each galaxy.
    $\theta_{i}$ is the angle between the $z$ axis and the vector of
    the angular momentum of a disk.
    $\phi_{i}$ is the azimuthal angle measured from $x$ axis to
     the projection of the angular momentum vector of a disk on
    to $xy$ plane. 
In the present study, we show the results of only one
model with $\theta_{1}$ = 30.0, $\theta_{2}$ = $-180.0$,
$\phi_{1}$ = 0.0,  $\phi_{2}$ = 0.0:
This model describes a nearly 
prograde-retrograde merger.
The results of the models with variously
different  $\theta_{i}$, $\phi_{i}$, $r_{\rm p}$, $e$, 
and $r_{\rm in}$ will be  described in our future papers (Bekki \& Shioya
2000c). 
The time when the progenitor disks merge completely and reach  the
dynamical equilibrium is less than 15.0 
in our units for the present model
($\sim$ 2.12 Gyr).

    Star formation
     is modeled by converting  the collisional
    gas particles
    into  collisionless new stellar particles according to 
    the Schmidt law (Schmidt 1959)
    with exponent $\gamma$ = 2.0 (1.0  $ < $  $\gamma$
      $ < $ 2.0, \cite{ken89}).
Chemical enrichment through star formation during galaxy merging
is assumed to proceed both locally and instantaneously in the present study
The fraction of gas returned to interstellar medium in each stellar particle
and the chemical yield
are 0.3 and 0.02, respectively.
Initial metallicity $Z_{\ast}$ for each stellar and gaseous
particle in a given galactic radius  $R$  (kpc) from the center
of a disk is given
according to 
the observed relation $Z_{\ast} = 0.06 \times {10}^{-0.197 \times (
R/3.5)}$ 
of typical late-type disk galaxies (e.g., Zaritsky, Kennicutt, \& Huchra 1994).
 For calculating the SED
that is not modified  by dust extinction for a merger,
we use the
spectral library GISSEL96 which is the  latest version of  Bruzual \& Charlot
(1993).
For deriving the SED modified by dust extinction,
we use our original  ray tracing method that
can be applied to  N-body simulations. The details of this
method and advantages and disadvantages in using this method 
are given in Bekki \& Shioya (2000a).
   All the calculations related to 
the above dynamical evolution  including the dissipative
dynamics, star formation, and gravitational interaction between collisionless
and collisional component 
 have been carried out on the GRAPE board
   (\cite{sug90})
   at Astronomical Institute of Tohoku University.
   The parameter of gravitational softening is set to be fixed at 0.03  
   in all the simulations. The time integration of the equation of motion
   is performed by using 2-order
   leap-flog method. Energy and angular momentum  are conserved
within 1 percent accuracy in a test collisionless merger simulation.

Based on the above numerical model, we investigate mainly
the SED and the photometric evolution of major mergers
and thereby provide some implications on the origin of
ULIRGs.  Before describing our numerical results,
we should point out that total infrared luminosity
($L_{\rm IR}$) of a major merger depends strongly on
the initial disk mass of the merger.
Accordingly, the $L_{\rm IR}$ of a major merger during starburst
can exceed $\sim$ $10^{12}$ $L_{\odot}$ that is a defined value
for ULIRGs (Sanders \& Mirabel 1996), only if
the disk mass of the merger exceeds a certain value (for
a given orbital configuration and gas mass fraction in major merging).
A possible dependence of $L_{\rm IR}$ on the initial disk mass of
a major merger is described in detail in Appendix A.
In  the present model with the initial disk mass of 
6.0 $\times$ $10^{10}$ $ \rm M_{\odot}$,
the maximum $L_{\rm IR}$ is $\sim$ 5.9 $\times$
$10^{11}$ $L_{\odot}$,
which is appreciably smaller than the defined value of ULIRGs.
Accordingly, the present merger model with structural and kinematical properties
nearly the same as those of the Milkey
way does not
show the evolution of ULIRG  
in a strict sense.
We however consider that it is not unreasonable
for us  to discuss extensively the origin and the nature
of ULIRGs by using the present model.
This is because some changes of merger parameters
(e.g., gas mass fraction, initial disk mass, and orbital
configurations of mergers)
can increase the $L_{\rm IR}$ a factor of 2
and because the defined value of $\sim$ $10^{12}$ $L_{\odot}$
itself does not probably have essentially important physical meanings.
Bekki \& Shioya (2000a) have already demonstrated that
if the gas mass fraction of initial disks of a merger
is rather large (0.5, which is an unrealistic value for the present
disk galaxies but not so an unrealistic one for the high
redshift dusty starburst galaxies such as faint submillimeter  sources
discovered by Smail et al. 1997),
the maximum infrared luminosity exceeds  
$10^{12}$ $L_{\odot}$.
In the present study,
we described the model with $f_{\rm g}$ = 0.2
and a nearly prograde-retrograde orbital
configuration as a fiducial one showing a typical behavior
of dusty major mergers that can form ULIRGs. 
The essentially important results on photometric evolution
of dusty starburst mergers do not depend so strongly on
orbital configurations and internal structure in mergers,
though the epoch and the strength of starbursts depend on
these parameters.
The dependence of photometric evolution on merger parameters
will be described in our future papers (Bekki \& Shioya 2000c).

\placefigure{fig-1}
\placefigure{fig-2}
\placefigure{fig-3}
\placefigure{fig-4}
\placefigure{fig-5}
\placefigure{fig-6}

\section{Result}

\subsection{Morphological evolution and star formation history}

Figures 1, 2, 3, and 4 describe the time evolution of global
morphology of the present merger model for each of three 
components, bulge, old disk, and gas disk (and new stars). 
>From now on, for convenience, the time $T$ represents the time that has
elapsed since the two disks begin to merge.
Seven fundamental properties of morphological evolution
of the less inclined and nearly prograde-retrograde merger 
with bulges are described as follows.
Firstly, a strong stellar bar is  not developed even in the disk orbiting
in a prograde sense for the first encounter of the merger 
(0.56 $<$ $T$ $<$ 0.85 Gyr). 
This is  because the central massive bulges of the two disks
can prevent the growth of non-axisymmetric perturbation (triggered
by tidal interaction of the merger).
Secondly, as galaxy merging proceeds, 
the two disks are strongly disturbed
to form a long tidal tail in the disk orbiting in a prograde sense
at 0.56 $<$ $T$ $<$ 0.71  Gyr.
A strong 
tidal tail is not developed in the disk orbiting in a retrograde sense.
This  one long tidal arm is characteristics of prograde-retrograde
mergers. 
Thirdly, the size of one disk orbiting in a prograde sense becomes  
smaller owing to the efficient tidal stripping of disk components
whereas that of the other disk orbiting in a retrograde
sense does not change so greatly for  1.12 $<$ $T$ $<$ 1.48 Gyr.
As a natural result of this, the major merger $could$  be observationally
identified as a minor merger with the mass ratio of the two disks rather
small ($m_2$ $<$ 0.1) or as unequal-mass one ($m_2$ $\sim$ 0.3)
during violent interaction of the major merger (corresponding
to  1.12 $<$ $T$ $<$ 1.48 Gyr in the present study). 
Fourthly, the tidal tail formed in the first encounter
becomes less discernible when
the two disks finally sink into the center of massive dark halos
owing to dynamical friction during merging and consequently
are completely destroyed by violent relaxation of
galaxy merging (1.41 $<$ $T$ $<$ 1.83  Gyr).  
Fifthly, an elliptical galaxy formed in the present very dissipative
and less inclined merger model has a rather flattened shape.
Sixthly, old stellar components initially located within  disks
and bulges are more diffusely distributed than new stars in
the merger remnant, primarily because new stars that were previously
gaseous components experience much more amount of gaseous dissipation
during merging.
Seventhly, these results are broadly consistent with those of previous
dissipative merger models such as Mihos \& Hernquist (1996).

 During violent major merging, interstellar gas is very
efficiently transferred to the central region of the two disks
owing to gaseous dissipation of colliding gas clouds
and gravitational torque.
Gas accumulated in the central region of the merger is
then consumed by massive starburst and consequently
converted into new stars.
Figure 5 describes the total gas mass accumulated
within the central 100, 200, and 500 pc for each of the
merging disks.
During the first tidal encounter of the merger 
(0.56  $<$ $T$ $<$ 1.12  Gyr),
tidal force transforms the disk orbiting in a prograde sense
more dramatically  and furthermore changes the mass distribution of  
the disk.  Consequently  angular momentum redistribution
of gas and stars proceeds in the disk and some fraction of gas
can be accumulated into the central region of the disk
(and some fraction of gas is 
tidally stripped away from the disk as a natural result of angular momentum
redistribution).
On the other hand,  the disk orbiting in a retrograde sense
does not suffer so severely from such tidal effects, and accordingly
the mass distribution is not greatly changed for 0.56  $<$ $T$ $<$ 1.12  Gyr. 
Thus, as is shown in Figure 5, total gas mass accumulated
in the central region is larger for the disk
orbiting in a prograde sense for 0.56  $<$ $T$ $<$ 1.12  Gyr.
For the final violent relaxation phase (1.41  $<$ $T$ $<$ 1.69  Gyr), 
the total gas mass 
is not so largely different and the gas mass is appreciably larger
for the disk orbiting in a retrograde  sense (in particular,
for the central 100 pc region). 
These results imply that the difference in the central gas mass
between two disks in a merger can be more clearly observed
in the merger early phase when two cores are still well separated
(i.e., the epoch of early tidal encounter).
As is shown in Figure 6,
owing to the large central  bulge,
strong starbursts can occur only in the late phases of the merger 
when two disks are completely destroyed by violent relaxation 
to form the central high density core
of an elliptical galaxy.
The star formation rate becomes  the maximum  value
of $\sim$ 120  $M_{\odot}$ ${\rm yr}^{-1}$ at $T$ = 1.53 Gyr. 
The time scale within which the star formation rate
exceeds 100 $M_{\odot}$ ${\rm yr}^{-1}$ is rather short (less than
0.1 Gyr), which implies that the time scale for a gas-rich
merger to become an ULIRG is an order of 0.1 Gyr. 

\placefigure{fig-7}
\placefigure{fig-8}
\placefigure{fig-9}
\placefigure{fig-10}
\placefigure{fig-11}

\subsection{Spatial  distribution of gas and stars}

Absorption of stellar  light for stars in a galaxy
can be  modeled according
to the following reddening formulation (Black  1987; Mazzei et al. 1992);
\begin{equation}
E(B-V)=N(\rm H)/4.77 \times {10}^{21} {\rm cm}^{-2} 
\times ({\it Z}_{\rm g}/0.0 2) \; \; ,
\end{equation}
where $N(\rm H)$ and ${Z}_{\rm g}$ are
gaseous column density and gaseous metallicity, respectively.
If we  follow this reddening formulation,
stellar light from a star is more strongly affected by dusty gas for
the higher  density and the larger metallicity of gas surrounding the star.
Accordingly estimating  gas density and metallicity around each of
stars in a galaxy is very important for deriving the total amount of dust
extinction and thus the SED modified by dust extinction.
Thus  we here show the time evolution of 
gaseous density and metallicity around 
stars before describing the time evolution of SEDs of the 
present merger model.
Figure 7 gives the time evolution of the mean value
for gaseous density, metallicity, and the degree of dust extinction
around new stars in the present merger model.
In deriving the mean value for each time $T$ in this figure, we first find
gaseous particles located within 0.02 (in our units; corresponding
to 350 pc)  of each stellar particle and thereby
estimate gaseous density and metallicity around each stellar particle.
Next we calculate the product of gaseous density and metallicity
for each stellar particle
(This product  is proportional to E(B-V) in the above reddening formulation).
For convenience we call this product as the degree of extinction
in explaining Figure 7, 8, 9,  and 10.
Based on the above three values of each particle, 
we calculate the mean of these values at each time $T$ 
both for old stars initially located in disks and bulges
and for new ones  formed from gas. 
As is shown in Figure 7,  both the mean gaseous density and the
mean gaseous metallicity around stars are larger in new stars
than in old ones. As a natural result of this,
the mean of the degree of extinction is considerably larger
in new stars than in old ones during galaxy merging.
The reason for these dependences is essentially that
new stars are formed preferentially in the merger central region with
very high density gas and are more compactly distributed within
the central cores of the merger compared with old stars.
These results accordingly suggest that SEDs from new stars
are more dramatically modified by dust than those from
old stars during dynamical evolution of the merger.
Furthermore the difference in the mean of the degree of 
dust extinction between new stars and old ones becomes
 larger during massive starburst (1.48 $<$ $T$ $<$ 1.62 Gyr),
mainly because a large amount of gas is the most efficiently transferred
to the nuclear regions where young stars are located.
This result implies that since most of stellar light comes from
young stellar components during starburst, the merger
shows the largest degree of dust extinction during its starburst.

Figure 8 shows the age of each star and the gaseous metallicity
around the star for all new stars at $T$ = 1.53 Gyr corresponding
to the epoch when the merger
show  the maximum star formation rate of 
$\sim 120$ $M_{\odot}$ $\rm yr^{-1}$.
For a given  stellar age, gaseous density, gaseous metallicity,
the degree of extinction around stars are all very  widely distributed,
which reflect the fact that both chemical evolution
and local density enhancement of gas proceed in a very
inhomogeneous way during starburst of the merger.
The number excess of stars around the age of 0 Gyr 
in Figure 8 is due mainly to
the massive starburst around $T$ = 1.53 Gyr.
Furthermore,  a  larger number of new stars are  located in the region
with  larger  dust extinction (the upper
parts in Figure 8) for the age of $\sim$ 0 Gyr.
Figure 9 shows the radial dependence of gaseous density, gaseous metallicity,
mean age of new stars (estimated for each radial bin), and the mass
fraction of new stars to all stars (old stars and new ones)
at $T$ = 1.53 Gyr (the maximum starburst epoch).
As is shown in this figure,
both gaseous density and  gaseous metallicity
are  likely to be higher and  mean age of new stars
is  likely to be younger in the inner region of the merger.
This suggests that stars with younger ages can be more heavily
obscured by dust than those with older ages:
The dust extinction is very selective in the
sense that only very young components can be heavily obscured
by dust. 
The selective dust extinction processes are described in detail in later
sections.
The fraction of new stars is  likely to be larger in the 
inner region of the merger, which implies that
the SED of the inner region of the merger is more  
strongly affected by dust extinction.

Figure 10 and 11 show  clearly that the dust extinction
can be  very selective during starburst in the merger.
Figure 10 describes the dependence of mean gaseous density,
mean gaseous metallicity, and mean of the degree of extinction
around new stars on the mean age of new stars (with ages
smaller than 1.2 Gyr) at $T$ = 1.48 (the beginning of starburst),
1.53 (the maximum starburst), 1.55, 1.62, and 1.83 Gyr (the end of 
starburst).  
The top panel of Figure 10 demonstrates that  irrespectively
of time $T$, the mean gaseous density
around  new stars is  likely to be 
larger for stars with younger ages
and furthermore that the age dependence 
is more clearly seen for the epoch of stronger starburst
($T$ = 1.53 and 1.55 Gyr).
For example,  mean gas density for  
new stars with the ages ranging from 0 Gyr to 0.06 Gyr 
is about a factor of 5.5 times larger than 
that for  new stars  with the ages ranging from 1 Gyr to 1.06 Gyr
at $T$ = 1.53 Gyr.
Furthermore the middle panel of Figure 10
indicates that although the mean gaseous metallicity around new 
stars is  likely to be larger for new stars with younger ages,
the age dependence is more clearly seen for the end of starburst
($T$ = 1.83 Gyr). 
The reason for the clearer age dependence of
gaseous metallicity around new stars at $T$ = 1.83 Gyr is that
chemical evolution in the central region where young stellar components
are located proceeds to a larger degree in the later phase of galaxy merging. 
As a result of the above age dependences,
the mean of the degree of dust extinction is  likely to be
larger for new stars
with younger ages at all time $T$ (1.48, 1.53, 1.55, 1.62,
and 1.83 Gyr) and the age dependence of the degree of extinction
is clearer for the epoch of stronger starburst
(See the bottom panel in Figure 10).
For example,  the mean degree of dust extinction
for new stars with the ages ranging from 0 Gyr to 0.06 Gyr 
is about a factor of 4.8 times larger than 
that for  new stars  with the ages ranging from 1 Gyr to 1.06 Gyr
at $T$ = 1.53 Gyr.
The age-dependent dust extinction derived in Figure 10 can be
seen also in Figure 11 for very young stars with the ages less
than 0.12 Gyr.  These results in Figure 10 and 11
thus demonstrate that dust extinction for young starburst components
in major gas-rich mergers
is very selective in the sense that stars with younger  ages are  
more heavily obscured by dusty gas.
This selective extinction is suggested to be crucially important
for explaining spectroscopic properties 
of dusty starburst galaxies (Poggianti \& Wu 1999; Shioya \& Bekki 2000).
We discuss this importance of selective dust extinction
in the section of discussion (\S 4.2).

\placefigure{fig-12}
\placefigure{fig-13}
\placefigure{fig-14}
\placefigure{fig-15}
\placefigure{fig-16}
 
\subsection{Time evolution of dust temperature,  $A_{V}$,
and SEDs}

Photometric evolution of dusty starburst galaxies
is basically determined by the SEDs,  and the SEDs
furthermore depend on the age-metallicity distribution of
stellar populations, the dust temperature (represented by 
$T_{\rm dust}$ from now on), and the degree of
dust extinction ($A_{V}$ for  extinction in  $V$ band).
Therefore the time evolution of  $T_{\rm dust}$ and  $A_{V}$
is important for photometric evolution of dusty starburst galaxies,
in particular, in infrared bands.  
In the present model, $T_{\rm dust}$ is calculated 
for each of gaseous particles and $A_{V}$
can be estimated for each of the stellar ones
(Bekki \& Shioya 2000a).
The SED of a galaxy thus depends  strongly on 
the detailed distribution of  $T_{\rm dust}$ and  $A_{V}$.
Figure 12 shows the location of each of stellar particle (with
different ages) on an age-$A_{V}$ plane at $T$ = 1.53 Gyr.
As is shown in this figure,  the value of $A_{V}$ is rather diverse;  
ranging from $\sim$ 0.0 mag to $\sim$ 300 mag for stars with  ages
larger  than 5.0 $\times$  $10^7$ yr.
This result reflects  the fact that gaseous density
and metallicity around stars depend
on the location of each star 
(i.e., whether a star is in the central high density
region or in the outer low density one)
and are thus  variously different between
stars.   
For stars with the ages less than 5.0 $\times$ $10^7$ yr,
there is a tendency that younger stars have larger $A_{V}$.
This result thus confirms the selective dust extinction
that is already described before in this paper. 
These young starburst populations are mostly located in
the central high density gaseous region
and $A_{V}$ of a new star 
is controlled greatly by the surrounding gaseous density.
A new star with younger age ($<$ 5.0 $\times$ $10^7$ yr),
which is located in the rather inner part
of the central starburst region,
is more heavily obscured by high density
gas in the central starburst region
and consequently the $A_{V}$ of the star is larger.

Figure 13 shows the location of each of gaseous particles
on $Z-T_{\rm dust}$ plane, where $Z$ is gaseous  metallicity.
We can discernibly observe a 
weak tendency that a more metal-enriched gas has a larger
temperature (in particular, for the metallicity range of $Z$ $<$ 0.02),
though the dispersion in $T_{\rm dust}$
distribution is rather large for a given $Z$. 
The main reason for this weak dependence is 
that a more metal-enriched gaseous particle
is  likely to be located in the central region
where young massive stars emit 
a larger amount of  stellar light 
that can be absorbed by dusty gas.
The age dependence of  $A_{V}$  and the metallicity
dependence of $T_{\rm dust}$ derived in Figure 
12 and 13 can be seen at other $T$ during starburst (Bekki \& Shioya 2000c),
which implies that such dependences are essential ingredients
of dusty starburst major mergers.
These dependences are furthermore important for (mean) 
$A_{V}$ and $T_{\rm dust}$ of the merger at each time $T$. 
Figure 14 shows the time evolution of $A_{V}$ and
$T_{\rm dust}$ in the merger.
Both  $A_{V}$ and $T_{\rm dust}$ gradually increase 
with time and become maximum at the epoch of maximum starburst
($T$ = 1.53 Gyr).
After starburst, $A_{V}$ rapidly decreases with time,
mainly because gaseous density around new stars becomes considerably
smaller owing to rapid consumption of gas by starburst (See also
Figure 7). 
$T_{\rm dust}$  also becomes low after starburst
because most of young massive
stars emitting a large amount of stellar light
(that can be absorbed by dusty gas) died out.
The time evolution of the SED and the shape of the SED
in the present  merger model depends strongly  
on the time evolution of $A_{V}$ and $T_{\rm dust}$.


Figure 15  describes the SEDs of the merger which we derive
based on the mass distribution of stellar and gaseous component
and the age and metallicity distribution
of stellar populations of the merger
at each time $T$.
The infrared and
submillimeter luminosities  become larger during 1.41  $<$ $T$ $<$ 1.53 Gyr
in the present merger.
This is firstly because star formation
rate, which is closely associated
with the total amount of stellar light absorbed by
interstellar dust,
becomes considerable  higher owing to the efficient gas transfer to
the central region of the merger and secondly because
the  density of dusty gas becomes also
very high in the later phase of the merging
(1.41 $<$ $T$ $<$ 1.53 Gyr)
so that the gas can heavily obscure the strong starburst.
After the maximum starburst at $T$ = 1.53 Gyr,
both infrared  and submillimeter luminosities  rapidly decline 
(1.53 $<$ $T$ $<$ 1.83 Gyr).
This is principally because most of interstellar gas indispensable for
strong starburst and dust obscuration is rapidly 
consumed by star formation in the  merger till $T$ $\sim$ 1.83 Gyr. 
Thus the time evolution of SED of a merger depends strongly on
that of the star formation rate, which is basically controlled
by the dynamical evolution of the merger. 

Figure 16 describes the SED of the merger for  old stars,
new ones, and gas at the maximum starburst ($T$ = 1.53 Gyr)
in models with and without dust extinction.
One of advantages of the present numerical model
is that we can investigate the SED of each of stellar particles
and the dust reemission from each of gaseous one.
Although
stellar light from new stars dominates the SED 
in $UV$  band ($\le$ 0.3 $\mu \rm m$) for the model without  dust extinction,
the difference in the $UV$ flux between the two stellar
populations is appreciably smaller  for the model with dust extinction.
This is mainly because new stars  are located in 
the central region,  where there is a large amount of high density gas,
and consequently  are  more heavily  obscured
by dust than old ones (See also Figure 8). 
On the other hand,
the contribution of  old stars to the SED in optical and near-infrared
bands is appreciably  larger  compared with that of new stars
in the model with dust extinction,
though the shape of the SED  at optical and near-infrared
bands is not significantly
different between new stars and old ones 
in the model without dust extinction.
This  result suggests that photometric properties of
dusty starburst mergers (i.e., Luminous Infrared Galaxies,
LIRGs and ULIRGs)
in optical and near-infrared band are determined largely by old
stellar populations (old stars) owing to selective
dust extinction of new stars.

\placefigure{fig-17}
\placefigure{fig-18}
\placefigure{fig-19}
\placefigure{fig-20}

\subsection{Photometric evolution}
Figure 17 describes the time evolution of absolute magnitude
from ultraviolet  to near-infrared wavelength (in $U$, $B$, $V$, $R$,
$I$, and $K$ band) for the models with and without dust extinction.
Firstly we summarize the evolution
of the model without dust extinction.
Owing to the aging of old stellar populations,
the absolute magnitude of the merger in
most of bands becomes fainter as galaxy merging
proceeds. The absolute magnitude in $U$ band is very sensitive
to star formation rate and therefore appreciably
brightens  at $T$ $\sim$  0.85 Gyr
when the two disks in the merger first encounter and consequently
weak starburst occurs in the disk rotating in a prograde sense.    
During strong starburst (1.4 $<$ $T$ $<$ 1.6 Gyr),
the absolute magnitude of the merger rapidly brightens with time
because of increased stellar light from young massive stars.
Furthermore, 
the increase of luminosity  depends on wavelength
such that the magnitude difference between the epoch of prestarburst ($T$ $<$
1.4 Gyr) 
and that of starburst is larger for shorter wavelength
(e.g., $\sim$ 1.7 mag for $U$ band and $\sim$ 0.5 mag for $K$ one).
The $K$ band absolute magnitude during starburst becomes
brighter  than $-25.2$1 
that is the mean value of ULIRGs (Sanders \& Mirabel 1996). 
The absolute magnitude of the merger
rapidly becomes  faint  with time after starburst
because of the death of young massive stars.
The  difference in absolute magnitude between starburst and poststarburst
is 
also larger for shorter wavelength.

Next we summarize the evolution of the model with dust extinction.
As is shown in Figure 17, the time evolution of the absolute magnitude
of the merger (0 $\le$ $T$ $\le$ 2.3 Gyr)
becomes rather moderate compared with the model without
dust extinction for all bands, though the time evolution
is qualitatively similar to that of the model without  dust extinction.
This is principally because young massive stars, which are important
factor for the rapid increase and decrease in photometric properties
of major mergers, are heavily obscured by dusty gas. 
The difference in absolute magnitude between the two models
with and without dust extinction 
is larger for shorter wavelength during starburst,
which reflect the fact that stellar light with shorter wavelength
can be more greatly absorbed by dust. 
Furthermore, irrespectively of wavelength, the difference in
absolute magnitude between the two models is  likely to
be the largest at the epoch of maximum starburst ($T$ = 1.53 Gyr).
This is  because young and very luminous components
formed by starburst are the most heavily obscured by dust at 
the maximum starburst epoch
when a large amount of dusty gas is transferred 
to the central region to form very  high
density region around the compact starburst.
The $K$ band magnitude at $T$ = 1.53 Gyr is about $-24.3$  mag 
that is about 0.9 mag fainter  (less luminous) than the mean
$K$ band magnitude of ULIRGs (Sanders \& Mirabel 1996).
This is either  because the initial total mass  of the
host disk of a typical ULIRG is about 2 times
larger than that of the Galaxy adopted as a merger progenitor
disk in  the preset study or because the present model
underestimates the star formation rate of the merger
(e.g., owing to the smaller initial gas mass fraction).

Figure 18 describes the time evolution of global colors 
from optical  to near-infrared wavelength (in $B-V$, $V-I$, $I-K$ and $R-K$)
for the models with and without dust extinction.
By comparing the results of the model with dust extinction
with those of the model without  dust extinction,
we can clearly observe that the effect of dust on 
color evolution is basically similar to that on 
the above evolution of absolute magnitude: The time
variation of colors during galaxy merging becomes rather moderate
owing to dust extinction.
To be more specific, although the $B-V$ color  
becomes $\sim$ 0.4 mag bluer in the maximum starburst ($T$ = 1.53 Gyr)
than in the prestarburst ($T$ = 1.4 Gyr) for the model without
dust extinction, the color difference between the two epochs
is only 0.04 mag for the model with dust extinction.
This result implies that a major merger does not 
necessarily show very blue colors when it becomes
a LIRG (or an ULIRG) owing to very strong  dust extinction. 
We here point out  that the color evolution derived above
is based only on the present model with a certain 
orbital configuration. 
We accordingly  suggest that although
the derived essential dust effects on color evolution
could  be true for  LIRGs and ULIRGs, 
the $absolute$ $magnitude$  of the colors of LIRGs and ULIRGs can depend
on orbital configurations and internal structure of galaxy
mergers.
Bekki \& Shioya (2000b)
found that a nearly retrograde-retrograde merger
suffers the more  remarkable dust extinction of
stellar light than a prograde-prograde merger.
Since the colors of mergers depend also on the degree of
dust extinction,
LIRGs and ULIRGs formed by mergers with variously different
orbital configurations and internal structure probably
show different colors.

Figure 19 shows the time evolution of mid-infrared and far-infrared
fluxes at 25 $\mu$m, 60 $\mu$m, and 100 $\mu$m of the merger,
that of the ratio of 25 $\mu$m flux to 60 $\mu$m one 
($f_{25 \mu \rm m}/f_{60 \mu \rm m}$),
and that of the ratio of 60 $\mu$m flux to 100 $\mu$m  one 
($f_{60 \mu \rm m}/f_{100 \mu \rm m}$).
The present model does not include the effects
of small grains on SEDs and accordingly can not so precisely
estimate the infrared flux around the order of 10 $\mu$m. 
Here we thus point out that although the following evolution
of the fluxes could be qualitatively similar to the evolution
observed in LIRGs and ULIRGs, the absolute values
of the fluxes can be largely different from observational ones. 
The 25 $\mu$m,  60 $\mu$m,
and 100 $\mu$m  fluxes are rather sensitive to star formation rate and
dust temperature and therefore depend strongly on
star formation histories and the degree of central gas concentration in
major mergers.
The first remarkable increase of the fluxes (in particular,
for 25 $\mu$m and 60 $\mu$m) corresponds to the epoch
of the first encounter of two disks of the merger ($T$ = 0.85 Gyr).
The second and the largest increase at $T$ = 1.53 Gyr
is due to the maximum starburst heavily obscured by dusty gas.
The increase of these fluxes is associated closely
with the increase of dust temperature and the fraction
of young stars obscured by dust.
As is shown in Figure 19, the flux ratio of 
$f_{60 \mu \rm m}/f_{100 \mu \rm m}$ 
becomes considerably  larger ($\sim$ 0.4) during starburst
compared with the initial value ($\sim$ 0.04).
This result implies that mergers in the merger late phase
when two cores become very close (less than a few kpc)
are more likely to show larger  
$f_{60 \mu \rm m}/f_{100 \mu \rm m}$.
The time evolution of the flux ratio of
$f_{25 \mu \rm m}/f_{60 \mu \rm m}$  
is rather moderate and furthermore the values of
$f_{25 \mu \rm m}/f_{60 \mu \rm m}$  
is less than 0.02 during merging.
The derived value is much smaller than that of the so-called
warm ULIRGs which are observationally suggested to
have AGNs (e.g., Sanders \& Mirabel 1996).
Thus our present study, 
which does not include the effects of small grains though,
confirms that dusty starburst alone can not contribute to
the large flux ratio of 
$f_{25 \mu \rm m}/f_{60 \mu \rm m}$  observed in warm ULIRGs.

Figure 20 gives the time evolution of $L_{\rm IR}$ and 
$L_{\rm IR}/L_{\rm B}$ in the present merger model.
Both  $L_{\rm IR}$ and $L_{\rm IR}/L_{\rm B}$  become
considerably larger during strong starburst,
because a larger number of young stars are 
created and then obscured
by dusty gas during starburst.
Although the  change of $B$ band magnitude 
is at most  0.3 mag during starburst (See Figure 17),
that of $L_{\rm IR}$ is $\sim$ an order of magnitude.
This result indicates that major mergers with similar luminosities
can show very different 
$L_{\rm IR}$ and furthermore that
$L_{\rm IR}$ reflects the dynamical stages of major galaxy merging.
Furthermore  the $L_{\rm IR}/L_{\rm B}$ during starburst becomes
larger than 10 that is similar to the order of $L_{\rm IR}/L_{\rm B}$
of LIRGs and ULIRGs.  
This result confirms that the rather large luminosity
ratio of $L_{\rm IR}/L_{\rm B}$  observed in LIRGs and ULIRGs
are associated closely with major galaxy merging.
Considering the results shown in Figure 19 and 20,
we can say that a merger with higher  $L_{\rm IR}/L_{\rm B}$ 
shows warmer $f_{60 \mu \rm m}/f_{100 \mu \rm m}$ 
and also that $f_{60 \mu \rm m}/f_{100 \mu \rm m}$
increases with the increase of $L_{\rm IR}$.
These derived correlations are consistent reasonably well with
the observational ones (e.g., Sandered \& Mirabel 1996).

\placefigure{fig-21}
\placefigure{fig-22}
\placefigure{fig-23}

\subsection{Two-dimensional  distribution}
 Figure 21 shows the two-dimensional distribution (projected onto  the 
$x$-$y$ plane) of $K$ band surface brightness, $R-K$ color,
60 $\mu$m flux, and $A_{V}$ at the epoch of maximum starburst
($T$ = 1.53 Gyr).
The $K$ band surface brightness is higher in the inner region
of the merger and shows peculiar double-peak morphology
in the central region. 
The main  reason for this central peculiar morphology
is that the merger 
at this epoch has  non-axisymmetric structure (composed both old stars and
new ones) 
formed by dynamical relaxation of major merging. 
The difference in $K$ band surface brightness between the central
region ($\sim$ 2 kpc) and the outer one ($\sim$ 8kpc) is 
about 7 mag ${\rm arcsec}^{-2}$ at this epoch.
The distribution of $R-K$ color 
is more complicated  and more strongly disturbed
than that of $K$ band surface brightness,
which implies that the signature of major galaxy merging
can be more clearly observed in the $R-K$ color distribution
than the $K$ band surface brightness distribution. 
The reason for this peculiar morphology
of color distribution is that 
stellar light in $R$ band (rather than in $K$ band)
is strongly affected by dusty gas that is distributed in
a very irregular way owing to dynamical perturbation of the merger.
The merger at the very luminous infrared phase
shows the  redder color in the central region
than in the outer region,
which means that ULIRGs and LIRGs are more likely to 
have negative color gradients.
The reasons for the derived negative gradient
are (1) that the central starburst components are rather heavily
obscured by dust and (2) that initially the central stellar components
are more metal-rich owing to the bulge components. 
The absolute magnitude of the negative $R-K$ color gradient
is estimated to be
0.08 mag/kpc for the central 8.5 kpc of the merger at this epoch.

As is shown in Figure 21, 
the infrared flux at 60 $\mu$m is distributed rather preferentially
to the central region of the merger.
The peak of the  60 $\mu$m  flux coincides both with 
the peak of the $A_{V}$ distribution and with 
the (lower) peak of the $K$ band surface brightness one. 
This result means that the infrared emission from dusty gas
comes exclusively from the central compact high density
gaseous regions where very young starburst components are located.
The coincidence of the two  peaks (in $K$ band surface brightness
and $A_{V}$) is furthermore associated with the origin
of peculiar morphology of
$K$ band surface brightness distribution.
The distribution of $A_{V}$  is also very compact and shows 
the rather large negative radial gradient ($\sim$ 3.0 mag/kpc for the
central 2 kpc). 
Figure 21 furthermore shows that 
the $R-K$ color distribution
is more similar to the $A_{V}$ one
than the 
$K$ band surface brightness  one. 
This implies that the two-dimensional distributions
of dust extinction in luminous infrared
galaxies formed by major mergers
are important not only for the infrared luminosity
distribution but also for  the color distribution.

\subsection{Correlations between physical properties}

 LIRGs and ULIRGs are observed to 
show a number of fundamental correlations
between physical properties such as $L_{\rm IR}$,  $L_{\rm IR}/L_{\rm B}$, 
infrared colors (e.g., $f_{60 \mu m}/f_{100 \mu m}$),
star formation efficiency estimated from the ratio of
$L_{\rm IR}$ to $\rm H_2$ molecular gas, and core separation of two
merging disks (Sanders \& Mirabel 1996; 
Gao \& Solomon 1999;
Young  1999).
For example, Gao \& Solomon (1999) 
found a correlation between 
the CO (1-0) luminosity 
and the projected separation of merger nuclei 
in a sample of 50 LIRG mergers.
Such observed correlations are considered to contain  valuable
information on the formation and the evolution of luminous  infrared 
galaxies (Sanders \& Mirabel 1996)
and thus should be explained by any theories of galaxy 
evolution and formation.
Figure 22 gives correlations between the separation of two merging
disks (cores) and five physical properties of the merger at
15 time steps ($T$ = 0, 0.28, 0.56, 0.71, 0.85, 0.99,
1.12, 1.41, 1.48, 1.53, 1.55, 1.62, 1.69, 1.83, and 2.26 Gyr).
In order to estimate of the core separation at each time step,
we initially
place a stellar (collisionless) particle in the center of mass of each disk
and thereby investigate the orbit of each particle.
By measuring  the separation of the two particles 
at each time step,
we estimate the core separation ($R$) of the merger.
As is shown in Figure 22, 
total gas mass within the central 500 pc is more likely
to be  smaller for the smaller core separation.
The decrease of gas mass with the increase of the core separation
for 0.0 $<$  $\log R$ $<$ 0.5 in our units (corresponding to 0.56 $<$
$T$ $<$ 0.99 Gyr) reflects the fact that  
the two disks  separate from  with each other 
after the first encounter of the disks and before final coalescence. 
The reason of the decrease of gas mass with
the decrease of core separation (in particular for
0.5  $<$ $\log R$ $<$ $-1.5$ in Figure 22) is simply  that
as the major merging proceeds,
a larger amount of gas within the central 500 pc
is consumed by star formation.
Gao \& Solomon (1999) 
found a correlation between 
the CO (1-0) luminosity corresponding to
a measure of molecular mass $M(\rm H_2)$ and
the projected separation of merger nuclei 
(the indicator of merging stages) in a sample of 50 LIRG mergers.
Since most of interstellar gas with rather high gaseous density
is observationally demonstrated to become $\rm H_2$ gas,
we consider that it is reasonable to compare the present
results on gaseous distribution with the observational
ones on $\rm H_2$ gaseous distribution. 
Thus, although the present study does not investigate  
the time evolution of $M(\rm H_2)$ gas mass, 
it is not unreasonable for us to say
the present numerical model  at least quantitatively
reproduces the observed correlation.
This result  in Figure 22 implies that the observed correlation
can be understood in terms of dynamical evolution of major merging.

The star formation efficiency estimated
from the ratio of star formation rate ($SFR$) to  total gas mass within the
central 500 pc is larger for the smaller core separation for 
$ log R$ $>$ $-1.2$ 
 (corresponding to $T$ $<$ 1.53 Gyr , i.e., before the
epoch of maximum starburst).
The reason of this derived dependence is that
owing to  the adopted  density-dependent star formation rate (i.e.,
the Schmidt law), the star formation efficiency becomes larger
when the two cores approach with each other and  
the degree of the central gas accumulation becomes larger.
After the maximum starburst, the star formation efficiency rapidly
declines with time and thus with the core separation ($T$ $>$ 1.53 Gyr),
which means that there is no remarkable positive correlation
between core separation and star formation efficiency for mergers
after massive starburst.
Gao \& Solomon (1999) also found
an anti-correlation between $L_{\rm IR}/ \rm M(H_2)$ corresponding to
a   global measure 
of the star formation rate per unit gas mass and the projected separation 
of merging two cores in LIRGs.
Our results successfully reproduce this observed dependence, in particular,
for the strong starburst epoch that can be interpreted as a LIRG phase.
We suggest that the observed correlation can be also understood
in terms of gas accumulation processes of major galaxy mergers.
Here,
we again point out that the gas mass in our simulations 
is not for $\rm H_2$ molecular mass estimated from
CO luminosity 
but for  gas mass (i.e., mass for  all gas components).
Thus more quantitative  comparison of our theoretical results
with observations   
(e.g., those by Gao \& Solomon 1999)
requires high-resolution numerical simulations that can derive
the time evolution of $\rm H_2$  mass (e.g., transformation
from HI gas to $\rm H_2$ one) 
and that of spatial distribution of $\rm H_2$ gas. 
We plan a  future study to  include the formation of
molecular gas ($\rm H_2$) from atomic one (HI)
in numerical simulations
and thereby to investigate time evolution of
 $\rm H_2$ mass and spatial distribution of  $\rm H_2$.

The luminosity ratio of $L_{\rm IR}/L_{\rm B}$, the flux
ratio of $f_{60 \mu \rm m}/f_{100 \mu \rm m}$,
and the degree of dust extinction ($A_{V}$) 
are  likely to be larger for the smaller core separation.
The essential reason for the dependence
of $L_{\rm IR}/L_{\rm B}$ 
is that stellar light in $B$ band  
of young stellar populations
is more heavily absorbed by dusty gas  and reemitted in 
the infrared band and consequently becomes smaller 
for the smaller core separation. 
The main reason for the dependence
of $f_{60 \mu \rm m}/f_{100 \mu \rm m}$
is that the dust temperature is likely to increase with the decrease of 
core separation.
The reason for the dependence
of $A_{V}$ 
is that the degree of the central gaseous accumulation
becomes larger as major merging (and dynamical relaxation) proceeds.
We accordingly suggest that   
$L_{\rm IR}/L_{\rm B}$, $f_{60 \mu \rm m}/f_{100 \mu \rm m}$, and $A_{V}$
observed in LIRGs and ULIRGs clearly depend
on to what degree galaxy mergers are dynamically relaxed.
We also suggest that the diversity of $L_{\rm IR}/L_{\rm B}$
observed in LIRGs and ULIRGs (e.g., Sanders \& Mirabel 1996) 
can be understood in terms of the difference in merging stages
(or in the degree of dynamical relaxation).
Figure 22 furthermore indicates that 
although the total mass of stellar components in the merger
does not increase significantly during galaxy merging (only
a factor of 0.15 increase),
there is an order of magnitude difference in
$L_{\rm IR}/L_{\rm B}$ between the merger at different dynamical stages.
Such large difference can be seen also in 
$f_{60 \mu \rm m}/f_{100 \mu \rm m}$, though the $mean$ $A_V$
does not change so greatly with time and core separation.
These results are only true for the present merger model with
a certain orbital configuration of galaxy merging.
Accordingly, the diversity in initial physical conditions
of galaxy merging (e.g., gas mass fraction and orbital configurations)  
could furthermore cause  considerably
greater difference in $L_{\rm IR}/L_{\rm B}$
and $f_{60 \mu \rm m}/f_{100 \mu \rm m}$.  
Thus we suggest that $L_{\rm IR}/L_{\rm B}$ and
$f_{60 \mu m}/f_{100 \mu m}$ are rather diverse (more than an order
of magnitude difference) in galaxy mergers with
variously different merger stages.

Figure  23 shows the location  of the present merger model on
the $L_{\rm IR}$ $-$ $L_{\rm IR}/L_{\rm B}$ plane, the $L_{\rm IR}$ $-$ 
$f_{60 \mu \rm m}/f_{100 \mu \rm m}$ one,
and  $f_{60 \mu \rm m}/f_{100 \mu \rm m}$ $-$  
$SFR/M_{\rm g} (R<500\rm pc)$ one 
for $T$ = 1.48, 1.53, 1.55, 1.62, and 1.69 Gyr
when the merger shows the infrared luminosity larger than
$10^{11}$ $L_{\odot}$.  
As is shown in this figure, $L_{\rm IR}/L_{\rm B}$  and 
$f_{60 \mu \rm m}/f_{100 \mu \rm m}$  are likely to be larger
for larger  $L_{\rm IR}$.
The main reason for the $L_{\rm IR}$ dependence of $L_{\rm IR}/L_{\rm B}$
is that 
both  $L_{\rm IR}$ and
the  $L_{\rm IR}/L_{\rm B}$ increase  
with  the increase of star formation rate.
The main reason for the $L_{\rm IR}$ dependence of
$f_{60 \mu \rm m}/f_{100 \mu \rm m}$ 
is that 
both  $L_{\rm IR}$ and
the dust temperature (an important determinant
the evolution of $f_{60 \mu \rm m}/f_{100 \mu \rm m}$)   increase  
with  the increase of star formation rate.
The $L_{\rm IR}$ dependences  derived in Figure  23
are  qualitatively consistent
with observational results (e.g., Sanders \& Mirabel 1996).
Figure 23 furthermore demonstrates 
that the star formation efficiency is likely to be
larger for the larger $f_{60 \mu m}/f_{100 \mu m}$
and thus that the present merger model
can reproduce qualitatively the observed correlation
between $f_{60 \mu m}/f_{100 \mu m}$ and star formation efficiency
(e.g., Young  1999).

\section{Discussion}

\subsection{The importance of relative distribution
of dusty gas and young stellar populations}

It is generally considered that investigating theoretically the time evolution
of dusty starburst galaxies is important for better understanding
the nature and the origin of ULIRGs. 
It is particularly important to investigate
spectral energy distributions  (SEDs) 
of dusty starburst galaxies,
primarily because  galactic SEDs are considered
to be one of essentially important factors that can determine
the nature of dusty starburst galaxies. 
The following previous theoretical studies
have succeeded
in clarifying  
important dependences of SEDs on physical parameters
such as spatial distribution 
of dust, geometries of stellar and gaseous components,
and dust properties
in  dusty starburst galaxies. 
Witt et al. (1991) investigated 
transfer processes of stellar light for models with variously
different spherical geometries of dust 
with  special emphasis on 
the effects of scattered light  on photometric properties
of galaxies 
and SEDs. 
Witt \& Gordon (1996) investigated the radiative transfer
processes of a central stellar source surrounded by a spherical,
statistically homogeneous but clumpy two-phase scattering medium
and found that the structure 
of dusty medium 
can greatly affect the conversion of UV, optical, and near-infrared radiation
into thermal far-IR dust radiation in a dusty system.
By comparing the observed SEDs of 30 starburst galaxies with
theoretical radiative transfer models  of dusty systems,
Gordon, Calzetti, \&  Witt (1997) 
discussed the importance of geometry of stellar and gaseous components
in determining the SED of a dusty starburst galaxy.

We here suggest  that among the important factors for galactic SEDs,
the relative spatial distribution of  
gas and  stars  is a particularly 
important determinant for the evolution of 
SEDs and thus for the photometric evolution
in dusty starburst galaxies.
In particular, we stress that
the relative spatial distribution of dusty interstellar 
gas and young stars (i.e., starburst populations) is 
critically important,  because
most of stellar light during starburst comes
from young starburst components in galaxies.
As is demonstrated in the present study,
photometric evolution  such as the time evolution
of $R-K$ color, $L_{\rm IR}$, $f_{60 \mu m}$  
and $L_{\rm IR}/L_{\rm B}$, clearly reflects the time evolution
of the spatial distribution  of dusty gas and young stars. 
The difference in the relative distribution of gas and
young stars is also suggested to explain the diversity in SEDs
and $L_{\rm IR}/L_{\rm B}$ observed in ULIRGs (Bekki \& Shioya 2000b, c).
Then can we observationally confirm the suggested importance of
the relative distribution of gas and young stars in   
photometric evolution of ULIRGs ?
Recent high-resolution observational studies on HI and $\rm H_2$  gaseous 
distribution have already revealed the detailed spatial
distribution of interstellar gas (e.g., Hibbard \& Yun 1999). 
Accordingly it will be  possible for us to answer the above question
if ongoing and future observational
studies  can derive the detailed spatial distribution of star-forming
regions and thus that of young stellar populations 
for a starburst galaxy consisting both of old and young stars. 
Based on the Keck spectra for each of the projected region of 3C48
(which is a QSO at $z$ = 0.38 and also identified as an ULIRG with
dusty starburst),
Canalizo \& Stockton (1999) examined the nature of stellar populations
for each of the regions
and thereby derived spatial distribution of young  stellar population
for 3C48.
However, there are only a few observational studies 
which investigate in detail 
the two-dimensional distribution
of spectroscopic properties
and address the projected distribution of young stellar populations
for dusty galaxies.
Young starburst populations are 
suggested to be  relatively correctly traced by the
spatial distribution of H$\alpha$ regions without
being affected seriously by dust extinction (Kennicutt 1998). 
Furthermore we can derive the detailed two dimensional
distribution of H$\alpha$ regions
by using  the technique of integral-field spectroscopy (e.g.,
Chatzichristou 1998).
Accordingly we expect  that future (and ongoing) systematic 
spectroscopic studies with integral-field spectroscopy
(for spatial distribution of star-forming regions)
combined with  very high-resolution mapping of dusty interstellar gas
will assess the validity of the suggested importance of
the relative distribution of gas and young stars. 

\subsection{Selective dust extinction}

Recently Poggianti \& Wu (1999) have investigated in detail
spectroscopic properties of ULIRGs and 
found that some ULIRGs show
both strong Balmer absorption and relatively modest [OII] emission
(the so-called e(a) spectra). 
This peculiar spectroscopic properties of ULIRGs are also
found in galaxies in distant clusters of galaxies (Dressler et al. 1999;
Poggianti et al. 1999; Smail et al. 1999),
which implies that this spectra is not characteristic only  
for ULIRGs.
Poggianti et al. (1999) suggested that this 
e(a) spectra comes from the starburst  regions obscured by dust.
Although a growing number of observational results on
e(a) galaxies with possible dusty starburst have been accumulated,
there are only a few theoretical studies addressing
the formation and the evolution of these e(a) populations.
Poggianti \& Wu (1999) first discussed that if the dust extinction
for a stellar population decreases with the stellar age,
e(a) spectra can be obtained
for a starburst galaxy.
They furthermore suggested that if the location and thickness
of dust patches depend on the age of the embedded stellar populations
in a starburst galaxy,
the effects of the above selective dust extinction become remarkable
for the galaxy. 
Based on a one-zone chemophotometric evolution model
of dusty starburst galaxies, 
Shioya \& Bekki (2000) confirmed that 
if a young starburst population is
preferentially obscured by dust than old one in
a dusty starburst galaxy, the galaxy shows e(a) spectra. 
Then, when and how the proposed difference in the degree of
dust extinction between stellar populations with different ages
is possible during galaxy merging?

We here suggest that the difference of dust extinction
between the central region and the outer one in a  galaxy
is a main cause for the above age-dependent extinction.
To be more specific, since younger stellar populations
formed by secondary starburst in the central region
with a larger amount of dust interstellar gas
are more heavily obscured by dust than the outer old
components,
the age-dependent extinction can be achieved in mergers.
This radial dependence of dust extinction is considerably
reasonable and realistic, considering that secondary dusty starburst
occurs preferentially in the central part of a galaxy
owing to efficient inward gas transfer driven by non-axisymmetric 
structure (such as stellar bars) during  galaxy merging. 
The present numerical simulations 
on  gaseous and stellar distribution in merging
disk galaxies with dusty starburst furthermore 
have demonstrated that the central young stellar component formed
by nuclear starburst 
is more heavily obscured by dusty gas than  the outer old component
initially located in merger progenitor disks. 
We accordingly  suggest that  
galaxies with the central dusty starburst (e.g., ULIRGs)
are more likely to
show the selective dust extinction and thus have spectroscopic
properties characteristic for e(a) galaxies. 
Although we have not yet investigated spectroscopic properties
of ULIRGs formed by major merging in detail,
it is not unreasonable to claim that the e(a) spectral appearance
is due essentially to the selective dust extinction (i.e.,
age-dependent extinction)
in the central regions of gas-rich major mergers with strong
dusty starburst.
In our preliminary studies (Bekki \& Shioya 2000c),
a dusty starburst merger shows k+a/a+k spectra (i.e., the classical
"E+A" galaxy)  at the epoch of post-starburst
and e(a) spectra at the epoch of maximum starburst.
 
\section{Conclusion}

We have investigated in detail morphological, structural, and photometric
evolution of gas-rich major mergers with strong dusty starburst
in an explicitly self-consistent manner in order to explore
the nature of ULIRGs.
Main conclusions are summarized as follows.

 (1) A massive starburst triggered by dynamical processes of major galaxy
merging is found to be very heavily obscured by high density,
dusty gas  in a gas-rich merger. 
As a result of this, the  merger  
can become a very luminous
infrared galaxy  with $L_{\rm IR}$ of $\sim$ 5.9 $\times$ 
 $10^{11}$ $L_{\odot}$ 
at the maximum starburst epoch which corresponds to
the epoch of the formation of central high density core of an elliptical
galaxy.

 (2) Dust extinction of stellar populations in a galaxy merger with
large  infrared luminosity ($L_{\rm IR}$ $>$  $10^{11}$ $L_{\odot}$)
is selective in the sense that 
younger stellar populations are preferentially obscured by dust
than old ones.
This is  because younger populations are located
in the central region where a larger amount of dusty
interstellar gas can be transferred from the outer gas-rich 
regions in the merger.
This selective extinction provides physical basis for the origin
of spectroscopic properties of ULIRGs, such as e(a) spectra
recently revealed by Poggianti \& Wu (1999).

(3) Both $L_{\rm IR}$ and $L_{\rm IR}/L_{\rm B}$ increase as the star formation
rate increases  during starburst of the present merger model,
resulting in the positive  correlation between $L_{\rm IR}$ 
and $L_{\rm IR}/L_{\rm B}$ 
in the merger model.
The time evolution of the relative  
distribution of dusty gas and young starburst components
is crucially important for the time evolution of these properties.

(4) The dust temperature $T_{dust}$ and the flux ratio of
$f_{60 \mu \rm m}/f_{100 \mu \rm m}$
increase with the increase of the star formation rate.
This is  because a larger number of young stars
formed by massive starburst can heat the dusty interstellar gas
as the star formation rate becomes higher.

(5) Global colors (e.g., $R-K$) are redder in the central region
than in the outer one at the epoch of strong starburst in a
dusty major merger,
which implies that most  ULIRGs show negative color gradients.
Furthermore the central surface brightness (e.g.,  in $K$ band) 
becomes higher as the star formation rate becomes larger during
starburst primarily because both old and  young stellar populations
are more strongly concentrated in the nuclear region of the merger. 
Two-dimensional color distributions   seem to be more strongly
disturbed than  surface brightness maps. 
The peak of two dimensional
distribution of  $f_{60 \mu \rm m}$ flux
coincides  with that of $A_{V}$.
These results clearly suggest that spatial distribution
of gas and stars is one of important determinants
for the two-dimensional photometric properties of major mergers.

(6) The star formation efficiency
in a major merger
becomes higher  
as the separation of the two cores  becomes smaller,
which clearly reflects the fact that dynamical processes of
galaxy merging play an important role in
determining  the time evolution
of dusty starburst in the merger (e.g., the time evolution of
chemical components and local gaseous density in the starburst regions). 
Total gas mass, 
the degree of dust extinction ($A_{V}$), $T_{\rm dust}$, 
$L_{\rm IR}$, $L_{\rm IR}/L_{\rm B}$, and $f_{60 \mu \rm m}/f_{100 \mu \rm m}$
also depend strongly
on the separation of two cores of the merger.

(7) The star formation efficiency 
during starburst in a merger 
is found to correlate positively with the flux ratio
of $f_{60 \mu \rm m}/f_{100 \mu \rm m}$, which is consistent with
recent observational results by Young  (1999).
Furthermore $L_{\rm IR}/L_{\rm B}$ and 
$f_{60 \mu \rm m}/f_{100 \mu \rm m}$
are  found to be proportional to $L_{\rm IR}$ and to the
central gaseous density of the  merger,
which is also consistent with fundamental properties
observed in ULIRGs (e.g., Sanders \& Mirabel 1996).

We lastly stress that better understanding of dissipative and dissipationless
dynamics of galaxy merging furthermore leads us to  
clarify more thoroughly
the photometric properties 
of ULIRGs.

\acknowledgments

We are  grateful to the  referee Guy Worthey for valuable comments,
which contribute to improve the present paper.
Y.S. thanks the Japan Society for Promotion of Science (JSPS) 
Research Fellowships for Young Scientists.

\newpage

\appendix

\placefigure{fig-24}
\placefigure{fig-25}

\section{The expected mass dependence of $L_{\rm IR}$}

The expected mass (luminosity) dependence of  $L_{\rm IR}$ is described 
as follows.
The $L_{\rm IR}$ is observationally suggested
to be  proportional to star formation rate (Kennicutt 1998). 
Therefore, the mass dependence of $L_{\rm IR}$
depends on the mass dependence of (maximum) star 
formation rate. Here we define the maximum value of
$L_{\rm IR}$ derived in the present study as $L_{\rm IR}(0)$.
Based on $L_{\rm IR}(0)$ for the model with the mass of the merger progenitor
disk equal to  6.0 $\times$ $10^{10}$ $ \rm M_{\odot}$,
we predict the $L_{\rm IR}(M_d)$ for mergers with
different $M_{d}$, where $M_{d}$ is the initial
total mass of a disk in a merger. 
We also  define the total mass   of 
luminous  and dark matter and size of the progenitor as 
$M_t$ and $R_d$, respectively.
We consider here 
that gas mass in a disk is  proportional
 to $M_d$ for simplicity. 
The dynamical time scale represented by $T_{\rm dyn}$ is given as  
\begin{equation}
T_{\rm dyn} \propto R_d^{3/2} M_t^{-1/2}.
\end{equation}
Assuming the Freeman's law and the constant ratio of $R_d$ to
the scale length of exponential disk,
we derive
\begin{equation}
\Sigma \propto M_d R_d^{-2} \sim {\rm const.}
\end{equation}
We assume
that the degree of self-gravity of a galactic disk is described as 
\begin{equation}
M_t \propto M_d^{(1 - \beta)} \; \; ,
\end{equation} 
where $\beta$ is a parameter that controls
the degree of self-gravitating of the disk
and  should be determined by observational
studies.
 $L_{\rm IR}(M_d)$ is proportional to star formation rate
and thus can be scaled to $M_{d}/T_{\rm dyn} \times 
L_{\rm IR}(0)$ (in our units). 
Then we can derive 
\begin{equation}
L_{\rm IR}(M_d) \propto M_d^{3/4 - \beta/2} \; \; .
\end{equation}
If we assume that  $\beta$ is 0 (0 $\le$ $\beta$ $\le$ 0.6,
Saglia 1996), 
this relationship predicts that $L_{\rm IR}(M_d)$ becomes larger as 
$M_d$ increases.
By using the derived value of 
5.9 $\times$ $10^{11}$ $ \rm L_{\odot}$ for $ L_{\rm IR}(0) $,
we can expect the following $M_d$ dependence of
$L_{\rm IR}(M_d)$,
\begin{equation}
L_{\rm IR}(M_d) = 5.9 \times 10^{11} (\frac{M_d}{6.0 \times 10^{10}
M_{\odot}})^{3/4} L_{\odot} 
\end{equation}
The expected mass dependences  of $L_{\rm IR}$ 
for the cases  of $\beta$ = 0.0 and 0.6 are  shown
in Figure 24.

\section{Orbital configuration of galaxy merging}

In this paper, we show the results of only one merger model with a 
prograde-retrograde configuration.
Figure 25 shows clearly this  orbital configuration of merging two disk
galaxies.  Stellar and gaseous particles in the galaxy labeled as Galaxy 1 
orbit  the center of the
galaxy in a counter-clockwise whereas the galaxy itself orbits the center of
mass of the merger in a counter-clockwise. This galaxy is considered
to be rotating in a prograde sense for the merger,
because intrinsic spin vector of the galaxy is nearly the same as that of  the orbital one. 
In the galaxy labeled as Galaxy 2, on the other hand,
stellar and gaseous components orbit   
the center of the 
galaxy in a clockwise whereas 
the galaxy itself orbits the center of
mass of the merger in a counter-clockwise. 
This galaxy is considered
to be rotating in a retrograde  sense
because intrinsic spin of the galaxy goes in the opposite
direction of the orbital one.
The merger with this orbital configuration is referred to as a nearly
prograde-retrograde merger in the present paper.

\clearpage


\figcaption{
Mass distribution of old stars  projected
onto $x$-$y$ plane 
at each time  $T$ for the  present merger model.
The time (in units of Gyr)  indicated in the upper
left-hand corner represents the time  that has elapsed
since the two disks begin to merge. 
Here the  scale is given in our units (17.5 kpc) and 
each of the  12  frames measures 105 kpc (6 length units) 
on a
side.    
The particles in the lower left  galaxy (referred to  as  Galaxy 1) 
orbit in a counter-clockwise, whereas those in
the upper right  one (Galaxy 2) orbit in  a clockwise.
Both of the galaxies  are assumed to orbit in a counter-clockwise,
which means that this merger is a prograde-retrograde one.
The details of this orbital configuration are  given in Appendix B.
\label{fig-1}}

\figcaption{
The same as Figure 1 but for  projected
onto $x$-$z$ plane. 
\label{fig-2}}

\figcaption{
Mass distribution of gas and new stars  projected
onto $x$-$y$ plane 
at each time  $T$ for the  present merger model.
The time (in units of Gyr)  indicated in the upper
left-hand corner represents the time  that has elapsed
since the two disks begin to merge. 
Here the  scale is given in our units (17.5 kpc) and 
each of the  12  frames measures 105 kpc (6 length units) 
on a
side.    
\label{fig-3}}

\figcaption{
Mass distribution of gas and new stars  projected
onto $x$-$z$ plane 
at each time  $T$ for the  present merger model.
The time (in units of Gyr)  indicated in the upper
left-hand corner represents the time  that has elapsed
since the two disks begin to merge. 
Here the  scale is given in our units (17.5 kpc) and 
each of the  12  frames measures 105 kpc (6 length units) 
on a
side.    
\label{fig-4}}

\figcaption{
Time evolution of gas mass accumulated within the central
500 pc (top), 200 pc (middle), and 100 pc (bottom)
for each of merger progenitor disks. 
Blue and red lines represent the results of the disk galaxy
orbiting in a prograde sense (represented by Galaxy 1) 
and those of the disk  orbiting in a retrograde sense (Galaxy 2),
respectively. Total mass is given in our units
(6.0 $\times$ $10^{10}$ $ \rm M_{\odot}$).
Note that a large amount of gas (6.0  
$\times$ $10^{9}$ $ \rm M_{\odot}$ for the central 500 pc) is accumulated
at $T$ $\sim$  1.5 (Gyr) when the
strong starburst occurs. 
\label{fig-5}}

\figcaption{
Time evolution of star formation rate  
in units of $M_{\odot}$ ${\rm yr}^{-1}$.
\label{fig-6}}

\figcaption{
The time evolution of 
mean gaseous density (top), mean gaseous metallicity (middle),
and the mean degree of dust extinction (bottom) around stars 
for old stars (solid) and new ones (dotted).
Note that owing to the higher gaseous density and
the larger gaseous metallicity around new stars,
new stars show a considerably larger degree of dust extinction
during starburst than old ones.
\label{fig-7}}

\figcaption{
Gaseous density (top), gaseous metallicity (middle),
and the degree of dust extinction (bottom) around stars are
plotted against age for each of new stars (represented by crosses)
at $T$ = 1.53 Gyr corresponding to the maximum starburst epoch.
The ages of new stars range from nearly $\sim$ 0 to $\sim$ 
1.5 Gyr at this epoch. 
The density is given in our units  (1.12 $\times$ $10^{-2}$ 
$M_{\odot}$ ${\rm pc}^{-3}$).
If there are no gaseous particles around a star,
gaseous metallicity is set to  0 for the star.
Here the degree of dust extinction for a star
is measured from the product of the gas density 
and the gas metallicity around the star.
Note that all the above three quantities
are rather widely distributed for a given stellar age. 
\label{fig-8}}

\figcaption{
Radial ($R$) distribution of gaseous density (upper left, in log scale),
gaseous metallicity (upper right), mean age of new stars (lower left),
and mass fraction of new stars to total disk mass (lower right)
at $T$ = 1.53 Gyr (maximum starburst epoch). 
At this epoch, the merger has  not relaxed dynamically yet,
and therefore the highest density region does not necessarily coincide
with the center of mass of the merger. 
Note that 
the merger shows the highest gaseous density,
the largest gaseous metallicity, the youngest stellar age,
and the largest mass fraction of new stars
around $R$ = 0.2  in our units (350 pc).
This means that young stellar populations,
most of which are located in the central region of the merger,
are rather preferentially obscured by dusty gas. 
\label{fig-9}}

\figcaption{
Dependences of mean gaseous density (top),
mean gaseous metallicity (middle), and mean of the degree
of dust extinction (bottom) around young stars on
the ages of young stars (with the age less than 1.2 Gyr)
at $T$ = 1.48 (blue), 1.53 (red), 1.55 (black),
1.62 (green), and 1.83 (magenta). 
The density is given in our units  (1.12 $\times$ $10^{-2}$ 
$M_{\odot}$ ${\rm pc}^{-3}$).
Note that irrespectively of the time $T$,
younger stellar populations
are likely to show the larger degree of dust extinction
owing to the higher gaseous density around them.
Note also that the dependence of dust extinction
is more appreciably seen in the epoch of stronger starburst 
($T$ = 1.53 and 1.55 Gyr). 
\label{fig-10}}

\figcaption{
The same as Figure 10 but for very young stars with the ages
less than 0.12 Gyr. The age dependence of dust extinction
is more clearly seen in this figure than in Figure 10. 
\label{fig-11}}

\figcaption{
The location of each new star on age$-A_{V}$ plane
at $T$ = 1.53 Gyr. 
Here $A_{V}$  is estimated from
the SED of each of stellar particle.
Note that the $A_{V}$ ranges
from $\sim$ 0.0 to $\sim$ 300. This is due mainly to the
large difference in gaseous metallicity and density
around new stars (See also Figure 8).
Note also that there is a tendency 
for younger stars to have larger $A_{V}$. 
\label{fig-12}}

\figcaption{
The location of each gaseous particle on $Z-T_{\rm dust}$ plane
at $T$ = 1.53 Gyr,
where $Z$ is  gaseous metallicity.
Here $T_{\rm dust}$  is estimated from
the SED of each of gaseous  particle.
Note that a gaseous particle with larger metallicity
is likely to have larger $T_{\rm dust}$. 
This is mainly because a gaseous particle around very young stars
has larger metallicity owing to rapid chemical enrichment around the
stars and show higher temperature owing to strong stellar radiation
from the stars.
\label{fig-13}}

\figcaption{
The time evolution of $A_{V}$ (upper) and $T_{\rm dust}$ (lower)
of the present merger model. Here   $A_{V}$ and $T_{\rm dust}$
are estimated from the SED of the merger at each time step. 
\label{fig-14}}

\figcaption{
The spectral energy distributions (SEDs) of
the present merger model
at each time  $T$ (top panel for $T$ = 0, 0.28,
0.56, 0.71, and 0.85 Gyr; middle for $T$ = 0.99, 1.12, 1.41, and 1.48 Gyr;
bottom for $T$ = 1.55, 1.62, 1.63, 1.82, 2.26 Gyr).  
\label{fig-15}}

\figcaption{The SED for each of three components,
old stars (red), new stars (black), and gas (dark blue)
for the model with dust extinction (upper) and
that without dust extinction (lower).  
Note that although the SED of the old stellar components
is not so different between the two models,
that of the new ones is very different between the two.
Clearly, this is due to the selective dust extinction.
\label{fig-16}}

\figcaption{
The time evolution of absolute magnitude in
$K$ (sky blue), $I$ (magenta), $R$ (green),
$V$ (dark blue), $B$ (red), and $U$ band (black)
for the model with dust extinction (upper)
and without dust extinction (lower).
\label{fig-17}}

\figcaption{
The time evolution of colors in
$R-K$ (green),
$I-K$ (dark blue), $V-I$ (black), and $B-V$ (red)
for the model with dust extinction (upper)
and without dust extinction (lower).
\label{fig-18}}

\figcaption{
Upper panel: The time evolution of far-infrared flux  in
 25 $\mu$m (black),
60 $\mu$m (red), and 100 $\mu$m band (dark blue).
Lower panel: The time evolution of the ratio
of 25 $\mu$m flux to 60 $\mu$m one (black)
and that of 60 $\mu$m flux to 100 $\mu$m one (red).
\label{fig-19}}

\figcaption{
The time evolution of $L_{\rm IR}$ (upper) 
and $L_{\rm IR}/L_{\rm B}$ (lower). 
\label{fig-20}}

\figcaption{
 Two-dimensional distribution of $K$ band surface brightness
(upper left; in units of mag ${\rm arcsec}^{-2}$),
$R-K$ color (upper right; mag), 60 $\mu$m surface brightness 
(lower left; $10^{-27}$  Jy/${\rm arcsec}^{-2}$, and $A_{V}$
(lower right; mag)
projected onto $x$-$y$ plane at  
$T$ = 1.53 Gyr in the present merger model.
Each frame measures 17.5 kpc on a side and includes 400  bins (20 $\times$ 20).
We here estimate the
mean values of these four properties   for each bin based on
SEDs of stellar particles located within each bin.
Color contours of $A_{V}$ (in lower right panel)
are described only for the regions
with the  $A_{V}$ larger than 0.1 mag  
in order that we can more clearly show the detailed distribution
of $A_{V}$.  
\label{fig-21}}

\figcaption{
Total gas mass within the central 500 pc (top; represented by
$\rm M_{\rm g}(R<500pc)$), 
the ratio of star formation rate
to total gas mass within the central 500 pc (the second
from the top; SFR/$\rm M_{\rm g}(R<500pc)$
corresponding to star formation efficiency),
the ratio of infrared luminosity to $B$ band one
(the third from the top; $L_{\rm IR}/L_{\rm B}$),
the ratio of 60 $\mu$m flux  to 100 $\mu$m  one
(the second from the bottom; $\rm f_{60}/f_{100}$),
and the degree of dust extinction (bottom; $A_{V}$)
are plotted against the separation of two disk cores in
the present merger at each of 15 time steps.
Each cross represents the location of the merger
at each time step.
The crosses are connected with
each other  according to the time sequence of the merger
in order that we can  more clearly show the evolution of the merger
on these planes.
The merger with larger core separation is more likely to be 
in the earlier merger stage. 
\label{fig-22}}

\figcaption{
These three panels describe correlations
between physical properties of the present
merger model with  
$L_{\rm IR}$ exceeding  $10^{11}$ $L_{\odot}$ (corresponding
to the epoch of very large infrared luminosity; $T$ = 1.48,
1.53, 1.55, 1.62, and 1.69 Gyr).
Top panel: A correlation between infrared luminosity ($L_{\rm IR}$)
in units
of $10^{11}$ $L_{\odot}$ and $L_{\rm IR}/L_{\rm B}$.
Middle panel: A correlation between infrared luminosity ($L_{\rm IR}$)
and the ratio of 60 $\mu$m flux to 100 $\mu$m one ($\rm f_{60}/f_{100}$).
Bottom panel: A correlation between 
the ratio of 60 $\mu$m flux to 100 $\mu$m one ($\rm f_{60}/f_{100}$)
and 
the ratio of star formation rate
to total gas mass within the central 500 pc 
(SFR/$\rm M_{\rm g}(R<500pc)$
corresponding to star formation efficiency).
\label{fig-23}}

\figcaption{
The expected dependence of $L_{\rm IR}$ on the initial
disk mass ($M_{d}$) for the case of 
$L_{\rm IR} \propto {M_d}^{3/4}$ 
(solid line, corresponding to $\beta$ = 0)
and for the case of $L_{\rm IR} \propto {M_d}^{0.45}$
(dotted one, $\beta$ = 0.6).
The physical meaning of $\beta$
is given in the main manuscript. 
The luminosity required for a merger to become
an ULIRG ($\sim$  $10^{12}$ $L_{\odot}$) 
is  also superimposed   by thick  solid line.
\label{fig-24}}

\figcaption{
Initial orbital configuration of the present merger model. 
The arrows describe which direction each of the two galaxies
rotates: 
The particles in the lower left  galaxy (labeled  as  Galaxy 1) 
orbit in a counter-clockwise whereas those in
the upper right  one (Galaxy 2) orbit in  a clockwise.
The galaxy orbit during merging (for 0 $\le$ $T$ $\le$ 1.13 Gyr)
is given by a solid line for the Galaxy 1 and by a dotted one
for the Galaxy 2. Each frame  measures 175 kpc on a side. 
\label{fig-25}}

\end{document}